\documentstyle[aps,multicol,prb,epsf]{revtex}
\begin{document}
\draft
\title{Adiabatic transport in nanostructures}
\author{O. Entin-Wohlman and Amnon Aharony}
\address{School of Physics and Astronomy, Raymond and Beverly Sackler
Faculty of Exact Sciences, \\ Tel Aviv University, Tel Aviv 69978,
Israel\\ }
\author{Y. Levinson }
\address{ Department of Condensed Matter Physics, The Weizmann
Institute of Science, Rehovot 76100, Israel}

\date{\today}
\maketitle

\begin{abstract}
A confined system of non-interacting electrons, subject to the
combined effect of a time-dependent potential and different
external chemical-potentials, is considered. The current flowing
through such a system is obtained for arbitrary strengths of the
modulating potential, using the adiabatic approximation in an
iterative manner. A new formula is derived for the charge pumped
through an un-biased system (all external chemical potentials are
kept at the same value); It reproduces the Brouwer formula for a
two-terminal nanostructure. The formalism presented yields the
effect of the chemical potential bias  on the pumped charge on one
hand, and the modification of the Landauer formula (which gives
the current in response to a constant chemical-potential
difference) brought about by the modulating potential on the
other. Corrections to the adiabatic approximation are derived and
discussed.

\end{abstract}


\begin{multicols}{2}

\section{Introduction and summary}

The flow of a dc current in response to a slowly-varying
time-dependent potential operating on an {\it un-biased} system is
termed ``adiabatic charge pumping".  \cite{AG99} This phenomenon,
first considered in Ref. \onlinecite{He91}, has attracted recently
much theoretical
\cite{Br98,AA98,Z99,SAA00,S00,Ma00,Lev00,W00,Mo01,Sh01,Ma01,wei2}
and experimental \cite{Sw99,Po92,Kou91} interest. In general,
``adiabatic pumping" occurs when the charge transferred across a
boundary during a single period of a certain modulating potential
is independent of the modulation frequency. This process is
adiabatic in the sense that the periodic potential varies very
slowly in time,  such that its frequency $\omega$  is smaller than
any characteristic energy scale of the electrons.

Eighteen years ago, Thouless \cite{Th83} has shown, using the
adiabatic approximation, that the ground state of an infinite
one-dimensional (1D) system of non-interacting electrons subject
to a slowly moving periodic potential can support a dc current.
Later theoretical investigations of quantum pumping in confined
nanostructures have employed the result derived by Brouwer,
\cite{Br98} which gives the pumped charge in terms of the
time-dependent scattering matrix related to the modulating
potential. The derivation presented by Brouwer is based on the
analysis of Ref. \onlinecite{Bu94}, which, in turn, utilizes an
expansion in the amplitude of the modulating potential in the
context of time-dependent scattering theory. Nevertheless, it is
generally accepted that the formalism of B\"{u}ttiker {\it et al.}
\cite{Bu94}  and the resulting Brouwer formula are valid
\cite{avron} even for large amplitudes, as long as the adiabatic
approximation holds.

This paper is devoted, among other issues,  to the exploration of
this point. We consider a spatially-confined system of
non-interacting electrons, connected by leads (denoted by
$\alpha$) to electronic reservoirs which are kept at various
chemical potentials, $\mu_{\alpha}$. The system is also subject to
a slowly-varying periodic potential. We derive an expression for
the instantaneous current flowing in this system, allowing for
arbitrary strengths of the modulating potential, using
time-dependent scattering states. The formalism is based on an
iterative solution of those states, in the adiabatic
approximation, in which the temporal derivative of the scattering
potential (and the scattering states) is the small parameter. The
formal derivation is summarized in Sec. II, where we obtain the
instantaneous current in the lowest-order adiabatic approximation.
This current is averaged over a single period of the modulating
potential.  In this way we obtain the effect of the chemical
potential bias on the pumped charge on one hand, and the
modification of the Landauer formula caused by the modulating
potential on the other. We find a new expression for the charge
pumped through an un-biased system [see Eq. (\ref{intpump})
below], which is particularly  useful in cases where the
modulating potential operates on the entire nanostructure,
\cite{aa} and is spatially-dependent. The Appendix includes the
derivation of the next order correction to the current.

We investigate the lowest-order expression for the current in Sec.
III, confining ourselves for simplicity to a system connected to
two reservoirs. Our results there can be summarized as follow. In
a two-terminal structure, the current averaged over a single
period, $\tau$,  consists of two parts. The first, denoted by
$I_{\rm pump}$, flows even when the system is un-biased, (but is
modified by the presence of the chemical potential difference). To
the lowest order in the adiabatic approximation, that current
reads
\begin{eqnarray}
&&I_{\rm pump}=\frac{e}{4\pi}\oint \frac{dt}{\tau}\int dE\Bigl
(-\frac{\partial}{\partial E}(f_{\ell}(E)+f_{r}(E))\Bigr
)\nonumber\\
&&\times \sum_{m}\Bigl [ \langle rm|\dot{V}|rm\rangle -\langle\ell
m|\dot{V}|\ell m\rangle\Bigr ],\label{intpump}
\end{eqnarray}
where $\ell $ and $r$ denote the left and the right leads,
respectively, and $f_{\ell ,r}(E)\equiv (e^{(E-\mu_{\ell
,r})/k_{\rm B}{\rm T} }+1)^{-1}$ are the Fermi distributions in
the reservoirs connected to the left and right leads. In Eq.
(\ref{intpump}), $|\beta m\rangle \equiv \chi^{t}_{\beta m}$ is
the instantaneous scattering state at time $t$, excited by an
incoming wave in channel $m$ of lead $\beta$, and $\dot{V}$ is the
temporal derivative of the time-dependent scattering potential.
Interestingly enough, the matrix element appearing in Eq.
(\ref{intpump}) can be written in terms of the instantaneous
scattering matrix, reproducing the Brouwer formula, \cite{Br98}
and vindicating its use for arbitrary amplitudes.  The second part
of the current, denoted by $I_{\rm bias}$, flows only when the
system is biased, i.e., $\mu_{\ell}\neq \mu_{r}$. When this
current is integrated over a single period, $\tau$, of the
modulating potential, one obtains the Landauer formula, modified
by the modulated potential in two ways: (i) The transmission
coefficient is the instantaneous one, $T^{t}$, averaged over a
single period of the time-dependent potential; (ii) There appears
a correction to the Landauer expression, which is related to the
temporal derivative of the modulating potential and the ensuing
instantaneous scattering matrix. In the simplest case in which the
two leads are single-channel ones, $I_{\rm bias}$ takes a
particularly simple form,
\begin{eqnarray}
I_{\rm bias}&=&\frac{e}{\tau\pi}\oint dt\int
dE\nonumber\\
&&\times\Bigl
[T^{t}+\frac{1}{2}T^{t}\frac{d\psi^{t}}{dt}\frac{\partial}{\partial
E}\Bigr ]\Bigl (f_{\ell}(E)-f_{r}(E)\Bigr ),\label{landauer}
\end{eqnarray}
in which $\psi^{t}$ is the instantaneous Friedel phase (i.e., the
transmission phase) of the nanostructure. The first term in Eq.
(\ref{landauer}) yields the Landauer formula for the present case.
The second term there is a correction, which is discussed in Sec.
III, using a simple example.

\section{Time-dependent scattering theory in the adiabatic
approximation}

In the first part of this section we solve for the time-dependent
scattering states iteratively, using the adiabatic approximation.
We then use those scattering states in the second part, to obtain
the current. The formalism presented below borrows from the
derivations in Refs. \onlinecite{L00} and \onlinecite{oew01},
extended to include the effect of a time-dependent scattering
potential.

\subsection{Time-dependent scattering states}

We consider a ballistic nanostructure of arbitrary geometry, which
consists of a nanostructure connected to several electronic
reservoirs. This system is described by the Hamiltonian
\begin{eqnarray}
{\cal H}({\bf r},t)={\cal H}_{0}({\bf r})+V({\bf r},t),
\end{eqnarray}
where the scattering potential $V({\bf r},t)$ is assumed to be
confined in space, so that asymptotic behaviors of the scattering
solutions can be defined un-ambiguously. This confined region is
attached to leads, numbered by the index $\alpha$, and each lead
is connected to a reservoir having the chemical potential
$\mu_{\alpha}$. The Hamiltonian ${\cal H}_{0}$ consists of the
kinetic energy.  We use the adiabatic approximation, requiring
that the characteristic inverse time-constant, $1/\tau$, which
describes the time dependence of $V$, is smaller than any
characteristic energy scale of the electrons. For a simple
oscillatory potential, $\tau =2\pi/\omega$.

As in the usual scattering treatment, we denote the incoming wave
with energy $E$ in lead $\alpha$ by $w_{\alpha n}^{-}$, where $n$
is the transverse mode number. This wave is a solution of the free
Hamiltonian,
\begin{equation}
({\cal H}_0({\bf r})-E)w_{\alpha n}^{-}({\bf r})=0, \label{H0}
\end{equation}
and is normalized such that it carries a unit flux. The scattering
solution of the full Hamiltonian, excited by this incoming wave,
can be written in the form
\begin{eqnarray}
\Psi_{\alpha n}({\bf r},t)&=&e^{-iEt}\chi_{\alpha n}({\bf
r},t),\nonumber\\
\chi_{\alpha n}({\bf r},t)&=&w^{-}_{\alpha n}({\bf
r})+\tilde{\chi}_{\alpha n}({\bf r},t). \label{scatsol}
\end{eqnarray}
The time dependence of the scattered wave function,
$\tilde{\chi}_{\alpha n}({\bf r},t)$, is expected to have the same
characteristic time scale as $V$. For example, when the modulating
potential is oscillating in time, $\tilde \chi$ contains all
harmonics of the frequency $\omega$.

The scattering solution $\Psi_{\alpha n}$ should satisfy the
time-dependent Schr\"{o}dinger equation,
\begin{eqnarray}
i\frac{\partial\Psi_{\alpha n}({\bf r},t)}{\partial t}={\cal
H}({\bf r},t)\Psi_{\alpha n}({\bf r},t).\label{sch}
\end{eqnarray}
Inserting (\ref{scatsol}) into (\ref{sch}), using Eq. (\ref{H0}),
we find
\begin{eqnarray}
\Bigl (G^{t}(E)\Bigr )^{-1}\tilde{\chi}_{\alpha n}({\bf
r},t)=V({\bf r},t)w^{-}_{\alpha n}({\bf
r})-i\frac{\partial\tilde{\chi}_{\alpha n}({\bf r},t)}{\partial
t},\label{chitild}
\end{eqnarray}
where $G^{t}$ is the instantaneous Green function of the full
Hamiltonian, such that
\begin{eqnarray}
\Bigl (E-{\cal H}({\bf r},t)\Bigr )G^{t}(E;{\bf r},{\bf
r}')=\delta ({\bf r}'-{\bf r}).\label{green}
\end{eqnarray}

We now solve Eq. (\ref{chitild}), using the adiabatic
approximation: the temporal derivative appearing on the
right-hand-side of that equation is regarded as a small correction
(of order $1/\tau$), and the equation is solved iteratively. The
zeroth-order is just the instantaneous scattering solution, which
we denote by $\chi^{t}_{\alpha n}$,
\begin{eqnarray}
\chi_{\alpha n}^{t}({\bf r})=w^{-}_{\alpha n}({\bf r})+\int d{\bf
r}'G^{t}(E;{\bf r},{\bf r}')V({\bf r}',t)w^{-}_{\alpha n}({\bf
r}').\label{chit}
\end{eqnarray}
The instantaneous scattering state $\chi^{t}_{\alpha n}$ is the
solution of the instantaneous Schr\"{o}dinger equation, with
energy $E$,
\begin{eqnarray}
\Bigl (E-{\cal H}({\bf r},t)\Bigr )\chi^{t}_{\alpha n}({\bf r})=0.
\label{chiteq}
\end{eqnarray}
Turning back to Eq. (\ref{chitild}), one finds that to first order
in the time-derivative the scattering solution reads
\begin{eqnarray}
\chi_{\alpha n}({\bf r},t)=\chi_{\alpha n}^{t}({\bf r})-i\int
d{\bf r}'G^{t}(E;{\bf r},{\bf r}')\dot{\chi}^{t}_{\alpha n}({\bf
r}'),\label{sol}
\end{eqnarray}
in which $\dot{\chi}_{\alpha n}^{t}$ is the time-derivative of the
instantaneous scattering state. Hence, to first order in the
adiabatic approximation the time-dependent scattering states are
given entirely in terms of the {\it instantaneous} solutions
(namely, $\chi^{t}_{\alpha m}$ and $G^{t}(E;{\bf r},{\bf r}')$) of
the problem at hand. One notes that the adiabatic solution
(\ref{sol}) of the scattering state is analogous to Thouless
\cite{Th83} solution for the ground-state wave function in his
model. In the Appendix, we discuss the corrections to the lowest
order adiabatic approximation.

\subsection{The current}

Here we outline the derivation of the current in the scattering
states formalism, as developed, e.g., in Refs. \onlinecite{L00},
\onlinecite{oew01}, and \onlinecite{Lev98}. One writes the field
operator of the electron, $\hat{\Psi}({\bf r},t)$, in terms of the
scattering states as
\begin{eqnarray}
\hat{\Psi}({\bf r},t)=\int
\frac{dE}{2\pi}\sum_{\alpha}\hat{a}_{\alpha n
}(E)e^{-iEt}\chi_{\alpha n}({\bf r},t),
\end{eqnarray}
in which $\hat{a}_{\alpha n}$ destroys an electron incoming in
channel $n$ of lead $\alpha$. The thermal average of the latter
operators is given by the Fermi distributions of the various
reservoirs, such that
\begin{eqnarray}
\langle \hat{a}^{\dagger}_{\alpha n}(E)\hat{a}_{\alpha '
n'}(E')\rangle =2\pi\delta (E-E')\delta_{\alpha n,\alpha '
n'}f_{\alpha}(E),
\end{eqnarray}
where $f_{\alpha}(E)$ is the Fermi distribution in the reservoir
connected to the $\alpha$ lead. With these definitions, the
thermal average of the current density operator becomes
\begin{eqnarray}
\langle {\bf j}({\bf r},t)\rangle
&=&\frac{e}{m}{\Im}\int\frac{dE}{2\pi}\sum_{\alpha}f_{\alpha}(E)\chi^{\ast}_{\alpha
n }({\bf r},t)\frac{\partial\chi_{\alpha n}({\bf r},t)}{\partial
{\bf r}},
\end{eqnarray}
where $e$ stands for the negative electron charge.

It is convenient to evaluate this quantity when ${\bf r}$
approaches $\infty$ in lead $\beta$ (which will be denoted by
${\bf r}\rightarrow\infty\beta$), and then to integrate the
current density over the cross-section of that lead (noting that
the incoming and outgoing waves are normalized to carry a unit
flux). In so doing, we may take advantage of the asymptotic
properties of the instantaneous quantities $\chi^{t}_{\alpha n}$
and $G^{t}(E)$, as documented in Refs. \onlinecite{L00} and
\onlinecite{Lev98},
\begin{eqnarray}
&&G^{t}(E;{\bf r},{\bf r}')|_{{\bf
r}\rightarrow\infty\beta}=-i\sum_{m}w^{+}_{\beta m}({\bf
r})\chi^{t}_{\beta m}({\bf r}'),\nonumber\\
&&\chi^{t}_{\alpha n}({\bf r})|_{{\bf
r}\rightarrow\infty\beta}=\delta_{\alpha\beta}w^{-}_{\alpha n
}({\bf r})+\sum_{m}w^{+}_{\beta m}({\bf r})S^{t}_{\beta m,\alpha
n}. \label{asymp}
\end{eqnarray}
Here, $w^{+}_{\beta m}$ is the outgoing wave in channel $m$ of
lead $\beta$ and $S^{t}_{\beta m,\alpha n}$ is the matrix element
of the instantaneous scattering matrix. As a result, the current
flowing into lead $\beta$ is given by
\begin{eqnarray}
&&I_{\beta}(t)=e\int\frac{dE}{2\pi}\sum_{\alpha n}f_{\alpha}(E)\nonumber\\
&&\times\Biggl (\delta_{\alpha\beta}-\sum_{m}\Bigl [ |S^{t}_{\beta
m, \alpha n}|^{2}-2\Re \Bigl ( S^{t}_{\beta m, \alpha n
}U^{\ast}_{\beta m, \alpha n}\Bigr )\Bigr ] \Biggr
),\label{gencur}
\end{eqnarray}
with
\begin{eqnarray}
U_{\beta m,\alpha n}=\int d{\bf r}\chi^{t}_{\beta  m}({\bf r})
\dot{\chi}^{t}_{\alpha n}({\bf r}).\label{u}
\end{eqnarray}
The result (\ref{gencur}) for the time-dependent current entering
into lead $\beta$ of the nanostructure holds for a general biased
system, whose various terminals have different chemical potentials
(as long as the time-dependence of the periodic potential is slow
enough). It is therefore interesting to consider charge
conservation, using that result. Indeed, summing (\ref{gencur})
over all leads, we obtain
\begin{eqnarray}
\sum_{\beta}I_{\beta}(t)=e\int \frac{dE}{2\pi}\sum_{\alpha n
}f_{\alpha}(E)\frac{d}{dt}\int d{\bf r}|\chi^{t}_{\alpha n}({\bf
r})|^{2},\label{chacon}
\end{eqnarray}
which shows that when the total current entering the system,
$\sum_{\beta}I_{\beta}(t)$, is integrated over a single period of
the modulating potential, the result is zero, i.e., the charge per
period is conserved. In deriving the result (\ref{chacon}) we have
employed (i) the unitarity of the instantaneous scattering matrix,
$\sum_{\beta m }S^{t\ast}_{\beta m,\alpha n}S^{t}_{\beta m,\alpha
'n'}=\delta_{\alpha n,\alpha 'n'}$; and (ii) the following
property of the scattering matrix \cite{L00,Lev98}
\begin{eqnarray}
\sum_{\beta m}S^{t\ast}_{\beta m,\alpha n}\chi^{t}_{\beta m}({\bf
r})=\chi^{t\ast}_{\alpha n}({\bf r}).\label{sums}
\end{eqnarray}

Equation (\ref{gencur}) can be considered as a generalization of
the Landauer formula, extended to include the effect of a
time-dependent potential, in the adiabatic approximation. The new
ingredient is the quantity $U_{\alpha n,\beta m}$, Eq. (\ref{u}).
This quantity can be expressed in terms of the temporal derivative
of the scattering potential,
\begin{eqnarray}
U_{\beta m,\alpha n}=\int d{\bf r}\Bigl
(-\frac{\partial\chi^{t}_{\beta m}({\bf r})}{\partial E}\Bigr
)\dot{V}({\bf r},t)\chi^{t}_{\alpha n}({\bf r}).\label{uder}
\end{eqnarray}
To prove this, we take the temporal derivative of Eq.
(\ref{chiteq}), and use Eq. (\ref{green}), to obtain
\begin{eqnarray}
\dot{\chi}^{t}_{\alpha n}({\bf r})=\int d{\bf r}'G^{t}(E;{\bf
r},{\bf r}')\dot{V}({\bf r}',t)\chi^{t}_{\alpha n}({\bf
r}').\label{chitt}
\end{eqnarray}
We insert this expression into Eq. (\ref{u}), and carry out one of
the spatial integrations using
\begin{eqnarray}
\int d{\bf r}'G^{t}(E;{\bf r},{\bf r}')\chi^{t}_{\alpha n}({\bf
r}')=-\frac{\partial \chi^{t}_{\alpha n}({\bf r})}{\partial E},
\label{dchide}
\end{eqnarray}
which follows directly by differentiating Eq. (\ref{chiteq}) with
respect to the energy, and using Eq. (\ref{green}) and the
symmetry of the Green function $G^{t}(E;{\bf r},{\bf
r}')=G^{t}(E;{\bf r}',{\bf r})$. This produces the result
(\ref{uder}).

\section{The two-terminal system}

Let us now confine ourselves to a nanostructure connected to two
terminals, with left ($\ell$) and right ($r$) leads. Then we can
use Eq. (\ref{gencur}) to write the current entering the system
from the left terminal in the form
\begin{eqnarray}
I_{\ell}(t)&=&e\int\frac{dE}{2\pi}\sum_{nm}\Biggl [\Bigl (
f_{\ell}(E)-f_{r}(E)\Bigr )\Bigl (|S^{t}_{rm,\ell
n}|^{2}\nonumber\\
&+&\Re\Bigl (S^{t}_{\ell m,\ell n}U^{\ast}_{\ell m,\ell
n}-S^{t}_{\ell
m,rn}U^{\ast}_{\ell m,rn}\Bigr )\Bigr )\nonumber\\
&+&\Bigl (f_{\ell}(E)+f_{r}(E)\Bigr )\nonumber\\
&&\times\Re\Bigl (S^{t}_{\ell m,\ell n}U^{\ast}_{\ell m,\ell
n}+S^{t}_{\ell m,rn}U^{\ast}_{\ell m,rn}\Bigr )\Biggr ].
\end{eqnarray}
An analogous expression holds for $I_{r}(t)$. The net current
flowing in the system during a single period of the modulating
potential then consists of two parts,
\begin{eqnarray}
I=\oint \frac{dt}{\tau}\Bigl (I_{\ell}(t)-I_{r}(t)\Bigr )=I_{\rm
bias}+I_{\rm pump},
\end{eqnarray}
where the first, $I_{\rm bias}$, flows only when the system is
biased, whereas the second, $I_{\rm pump}$, is established by the
time-dependent potential (though it is affected by the chemical
potential difference, when the latter is applied). Using Eqs.
(\ref{sums}) and (\ref{uder}), the pumped current, $I_{\rm pump}$,
takes the form
\begin{eqnarray}
&&I_{\rm pump}=e\oint\frac{dt}{\tau}\int\frac{dE}{2\pi}\Bigl
(f_{\ell}(E)+f_{r}(E)\Bigr )\nonumber\\
&&\times\frac{1}{2}\sum_{m}\Biggl [-\frac{\partial}{\partial
E}\Bigl (\langle\chi^{t}_{\ell m}|\dot{V}|\chi^{t}_{\ell m}\rangle
-\langle\chi^{t}_{r m}|\dot{V}|\chi^{t}_{r m}\rangle\Bigr )\Biggr
]. \label{ipump}
\end{eqnarray}
For the biased current we find
\begin{eqnarray}
&&I_{\rm bias}=e\oint\frac{dt}{\tau}\int\frac{dE}{2\pi}\Bigl
(f_{\ell}(E)-f_{r}(E)\Bigr )\nonumber\\
&&\times\sum_{nm}\Biggl [2|S^{t}_{rm,\ell n}|^{2} +\Re\Bigl
(S^{t}_{\ell m,\ell n}U^{\ast}_{\ell m,\ell
n}-S^{t}_{\ell m,rn}U^{\ast}_{\ell m,rn}\nonumber\\
&-&S^{t}_{rm,\ell n}U^{\ast}_{rm,\ell
n}+S^{t}_{rm,rn}U^{\ast}_{rm,rn}\Bigr )\Biggr ].\label{ibias}
\end{eqnarray}

The pumped part of the current, $I_{\rm pump}$, can be written in
terms of the temporal derivatives of the instantaneous scattering
matrix. In order to vindicate this statement, we start from the
asymptotic form for $\chi_{\alpha n}^{t}$, Eq. (\ref{asymp}), for
the wave going from lead $\alpha$ into lead $\beta$. Noting that
\begin{equation}
w_{\alpha n}^{-}({\bf r} \rightarrow \infty,\beta)=w_{\alpha n}
^{-}\delta_{\alpha\beta}+ (1-\delta_{\alpha\beta})\sum_{m}w_{\beta
m}^{+},
\end{equation}
we conclude that
\begin{eqnarray}
S^{t}_{\beta m,\alpha n}&=&1-\delta_{\alpha\beta} -i \int d{\bf
r}\chi_{\beta m}^{t}({\bf r})V({\bf r},t)w^{-}_{\alpha n}({\bf
r}). \label{scat}
\end{eqnarray}
Differentiating this expression with respect to time and using
Eqs. (\ref{chit}) and (\ref{chitt}), yields
\begin{eqnarray}
i\dot{S}^{t}_{\beta m,\alpha n}=\int d{\bf r}\chi^{t}_{\beta m
}({\bf r})\dot{V}({\bf r},t)\chi^{t}_{\alpha n}({\bf
r}).\label{sdot}
\end{eqnarray}
Note that this expression is {\bf not} a matrix element. To turn
it into an expression involving matrix elements, we use Eq.
(\ref{sums}), by which
\begin{eqnarray}
i\dot{S}^{t}_{\beta m,\alpha n}&=&\int d{\bf r}\dot{V}({\bf
r},t)\chi^{t}_{\beta m}({\bf r})\nonumber\\
&&\times\sum_{\beta 'm' }S^{t}_{\beta 'm', \alpha
n}\chi^{t\ast}_{\beta 'm'}({\bf r}).
\end{eqnarray}
Then, multiplying  by $S^{t\ast}_{\beta m,\alpha n}$ and summing
over $\alpha$ and $n$ yields
\begin{eqnarray}
&&\sum_{\alpha n}\dot{S}^{t}_{\beta m,\alpha n}S^{t\ast}_{\beta m,
\alpha n}=-i\int d{\bf r}\chi^{t\ast}_{\beta m}({\bf
r})\dot{V}({\bf r},t)\chi^{t}_{\beta m}({\bf r}).\label{brouwer}
\end{eqnarray}
One notes that this identity makes the expression for $I_{\rm
pump}$, Eq. (\ref{ipump}) above, to be identical with the Brouwer
\cite{Br98} formula, in the case where the system is un-biased.

In the simplest case where each of the leads is a single-channel
one, the expression for the current takes a particularly simple
form. In this situation, the instantaneous scattering matrix
becomes a $2\times 2$ matrix, which can be parametrized (in the
absence of a magnetic field) as
\begin{eqnarray}
S^{t}=e^{i\psi^{t}} \left [
\begin{array}
{cc} \sqrt{R^{t}}e^{i\alpha^{t}}&i\sqrt{T^{t}}\\i
\sqrt{T^{t}}&\sqrt{R^{t}}e^{-i\alpha^{t}}
\end{array}\right ].\label{smatrix}
\end{eqnarray}
Here, $T^{t}$ and $R^{t}$ are the instantaneous transmission and
reflection, respectively. The reflection phase $\alpha^{t}$
describes the asymmetry of the nanostructure. A finite
time-dependent reflection phase is a necessary ingredient to
obtain the pumped current. Finally, $\psi^{t}$ is the transmission
(Friedel) phase. With the parametrization Eq. (\ref{smatrix}) one
finds, using Eqs. (\ref{sdot}) and (\ref{brouwer}),
\begin{eqnarray}
&&I=\frac{e}{2\pi}\oint\frac{dt}{\tau}\int dE\Biggl [\Bigl
(f_{\ell}(E)-f_{r}(E)\Bigr )\Bigl [2T^{t}\nonumber\\
&-&\frac{1}{2} \Bigl (\frac{\partial}{\partial E}\Bigl
((R^{t}-T^{t})\frac{d\psi^{t}}{dt}\Bigr )+\frac{d}{dt}\Bigl
((R^{t}-T^{t})\frac{\partial\psi^{t}}{\partial E}\Bigr )\Bigr
)\Bigr ]\nonumber\\
&+&\Bigl (f_{\ell}(E)+f_{r}(E)\Bigr )\frac{\partial}{\partial
E}\Bigl ( R^{t}\frac{d\alpha^{t}}{dt}\Bigr )\Biggr ].
\end{eqnarray}
The charge passing through the nanostructure during a single
period of the potential, $Q$, is then
\begin{eqnarray}
Q=Q_{\rm bias}+Q_{\rm pump}.
\end{eqnarray}
Here,
\begin{eqnarray}
Q_{\rm bias}&=&\frac{e}{2\pi}\oint dt\int dE\Biggl [\Bigl
(f_{\ell}(E)-f_{r}(E)\Bigr )2T^{t}\nonumber\\
&+& \frac{\partial}{\partial E}\Bigl (f_{\ell}(E)-f_{r}(E)\Bigr
)T^{t}\frac{d\psi^{t}}{dt}\Biggr ].
\end{eqnarray}
It is seen that the first term here is just the Landauer formula,
with the transmission coefficient averaged over the temporal
period. The second term forms a correction to this result, brought
about by the modulating potential. The pumped charge is given by
\begin{eqnarray}
Q_{\rm pump}&=&-\frac{e}{2\pi}\oint dt\int
dE\nonumber\\
&&\times\frac{\partial}{\partial E}\Bigl
(f_{\ell}(E)+f_{r}(E)\Bigr )R^{t}\frac{d\alpha^{t}}{dt},
\end{eqnarray}
(see Ref. \onlinecite{SAA00}), and it vanishes unless $\alpha^{t}$
is time-dependent.

\subsection{Example--the single-level quantum dot}

Clearly, a comparison between $Q_{\rm pump}$ and the second part
of $Q_{\rm bias}$ is called for. Consider for simplicity zero
temperature. Then, $Q_{\rm pump}$ is given by the values of
$R^{t}d\alpha^{t}/dt$ at $E=E_{\rm F}\pm \delta\mu /e$, where
$\delta\mu$ is the chemical potential difference, and $E_{\rm F}$
denotes the Fermi level. On the other hand, the second term in
$Q_{\rm bias}$ is given by $\sim\int
dE[-(\delta\mu/e)(\partial^{2}f/\partial
E^{2})T^{t}(d\psi^{t}/dt)]$, and hence should be much smaller than
the Landauer contribution, which is proportional to $\delta\mu
/e$. Nevertheless, it may be of interest to explore this term
experimentally, as it is related to the Friedel phase of the
nanostructure.

To further explore this point, we consider the following simple
example: a quantum dot, with a single localized level, coupled to
two ideal 1D leads. \cite{ng} Adopting the tight-binding
description, we model the two leads connecting the quantum dot to
the electronic reservoirs by 1D chains of sites, whose on-site
energies are assumed to vanish, and whose nearest-neighbor
transfer amplitudes are denoted by $-J$. Thus the energy of an
electron of wave vector $k$ moving on such a chain is
\begin{eqnarray}
E_{k}=-2J\cos ka,
\end{eqnarray}
where $a$ is the lattice constant. The localized level, of energy
$\epsilon_{0}$, is attached to the left-hand-side lead with matrix
element $-J_{\ell}$, and to the right-hand-side lead with matrix
element $-J_{r}$. The latter two quantities are assumed to vary
slowly in time, in a periodic way. Our formalism requires just the
knowledge of the instantaneous scattering matrix of the system.
For the case at hand,
\begin{eqnarray}
S^{t}= \left [
\begin{array}
{cc} -1+ \Bigl (\frac{J_{\ell}}{J}\Bigr )^{2}M_{k}
&\frac{J_{\ell}J_{r}}{J}M_{k}\\
\frac{J_{\ell}J_{r}}{J}M_{k}& -1+ \Bigl (\frac{J_{r}}{J}\Bigr
)^{2}M_{k}
\end{array}\right ],\label{sng}
\end{eqnarray}
with
\begin{eqnarray}
M_{k}=\frac{2iJ\sin
ka}{E_{k}-\epsilon_{0}+e^{ika}(J_{\ell}^{2}+J_{r}^{2})/J}.
\end{eqnarray}

\begin{figure}
\leavevmode \epsfclipon \epsfxsize=8truecm
\vbox{\epsfbox{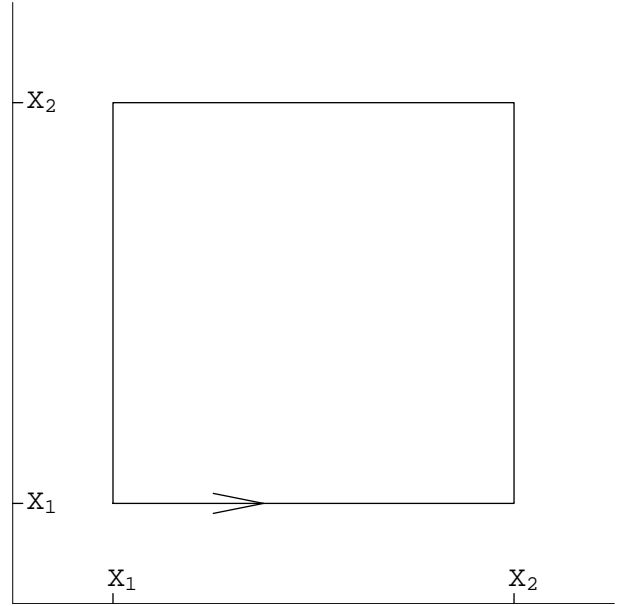}} \vspace{1cm} \caption{The periodic
temporal evolution of the parameters X$_{\ell}$ and X$_{r}$ in the
parameter plane. }\label{fig1}
\end{figure}

Let us consider first the pumped charge. At zero temperature, it
is given by
\begin{eqnarray}
&&Q_{\rm pump}=\frac{e}{2\pi}\oint dt R^{t}\frac{d \alpha^{t}}{d
t}\nonumber\\
&=&\frac{e}{2\pi}\oint dt \frac{\sin
ka}{|E_{k}-\epsilon_{0}+e^{ika}({\rm X}_{\ell}+{\rm
X}_{r})|^{2}}\nonumber\\
&&\times \Bigl [(\dot{\rm X}_{\ell}-\dot{\rm X}_{r})
(\epsilon_{0}-E_{k})+E_{k}({\rm X}_{r}\dot{\rm X}_{\ell}-{\rm
X}_{\ell} \dot{\rm X}_{r})\Bigr ],
\end{eqnarray}
in which energies are measure in units of $J$, and
\begin{eqnarray}
{\rm X}_{\ell}\equiv\Bigl (\frac{J_{\ell}}{J}\Bigr )^{2},\ \ {\rm
X}_{r}\equiv\Bigl (\frac{J_{r}}{J}\Bigr )^{2},
\end{eqnarray}
are the time-dependent parameters of the system. Note that these
two parameters can be thought of as the `contact conductances' of
the quantum dot. Now imagine those to vary in time as follows:
Initially, both are equal to X$_{1}$. Then X$_{\ell}$ is increased
linearly in time until it reaches the value X$_{2}$, while X$_{r}$
is being held fixed at the value X$_{1}$. From that point,
X$_{\ell}$ is held fixed, while X$_{r}$ increases linearly to the
value X$_{2}$, and so on, see Fig. \ref{fig1}.

\vspace{1cm}

\begin{figure}
\leavevmode \epsfclipon \epsfxsize=8truecm
\vbox{\epsfbox{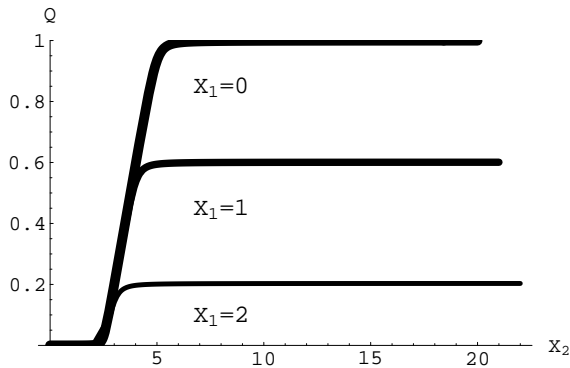}} \vspace{1cm}\caption{The pumped charge
(in units of $e$) as function of X$_{2}$, for several values of
X$_{1}$, indicated on the figure. Here $ka=\pi -0.1$, and
$\epsilon_{0}=-3$.}\label{fig2}
\end{figure}

It is quite straightforward to find $Q_{\rm pump}$ for such a
cycle. One obtains
\begin{eqnarray}
Q_{\rm pump}=\frac{e}{\pi}\int_{{\rm X}_{1}}^{{\rm X}_{2}}d{\rm X}
\Bigl (F({\rm X};{\rm X}_{1})-F({\rm X};{\rm X}_{2})\Bigr ),
\end{eqnarray}
with
\begin{eqnarray}
F({\rm X};{\rm Z})=\frac{\sin ka (\epsilon_{0}-E_{k}+E_{k}{\rm
Z})}{|E_{k}-\epsilon_{0}+e^{ika}({\rm X}+{\rm Z})|^{2}}.
\end{eqnarray}
The resulting charge differs significantly from zero, and
approaches unity (in units of $e$) as long as the line of maximal
transmission in the X$_{\ell}$-X$_{r}$-plane is well within the
closed orbit forming the period. \cite{Lev00} In the present
model, that line is given by
\begin{eqnarray}
{\rm X}_{\ell}+{\rm X}_{r}=2\frac{E_{k}-\epsilon_{0}}{E_{k}}.
\end{eqnarray}
We will elaborate on this point, which is exemplified in Fig.
\ref{fig2}, in a future publication. For the parameters used to
produce the results shown in Fig. \ref{fig2}, for X$_{1}=0$ the
maximal transmission line is contained within the pumping contour
(see Fig. \ref{fig1}), while for its higher values the pumping
contour shifts away from the maximal transmission line.

\vspace{1cm}

Let us next consider the time-average that appears in the second
term of the biased current, i.e., $\oint T^{t}d\psi^{t}/d t$. In
our example, this quantity becomes
\begin{eqnarray}
&&\oint dt T^{t}\frac{d\psi^{t}}{dt}=\nonumber\\
&-&\oint dt \frac{4\sin^{3}ka (E_{k}-\epsilon_{0}){\rm
X}_{\ell}{\rm X}_{r}(\dot{\rm X}_{\ell}+\dot{\rm
X}_{r})}{|E_{k}-\epsilon_{0}+e^{ika}({\rm X}_{\ell}+{\rm
X}_{r})|^{4}}.
\end{eqnarray}
On a symmetric periodical curve like that presented in Fig.
\ref{fig1} this integral vanishes. Hence, in such a symmetric
configuration, there will be no deviation from the Landauer
formula due to the temporal variation of the Friedel phase. One
may conclude from this example that when the pumping orbit in the
plane of the time-dependent parameters is not so simple as the one
considered above, there might be a small correction, related to
the asymmetry of the orbit.

\section{Concluding remarks}

The motivation of this work is mainly to examine the validity of
the adiabatic approximation in calculating the charge pumped
quantum-mechanically through a confined system of non-interacting
electrons, and to study the effects of a constant bias. To this
end, we have derived the time-dependent current generated when the
system is subject to a slowly-changing modulating potential, in
addition to being connected to reservoirs of different chemical
potentials. We have found that the widely-used Brouwer formula
which gives the pumped charge in terms of the instantaneous
scattering matrix is valid for arbitrary amplitudes of the
modulating potential, as long as the lowest-order adiabatic
approximation can be employed.

In the process, we have obtained an alternative expression for the
pumped charge, Eq. (\ref{intpump}), which gives it in terms of
matrix elements of the temporal derivative of the potential
between the instantaneous scattering states. We have also derived
and analyzed the effects of the modulating potential on the biased
current, and showed them to be quite small. Our formalism allows
for the possibility to obtain systematically the corrections to
the lowest-order adiabatic approximation.

\acknowledgements

We thank P. W\"{o}lfle for helpful conversations. This research
was carried out in a center of excellence supported by the Israel
Science Foundation, and was supported in part by the National
Science Foundation under Grant No. PHY99-07949 (at the ITP, Santa
Barbara), and by the Albert Einstein Minerva Center for
Theoretical Physics at the Weizmann Institute of Science.

\appendix

\section{Corrections to the adiabatic approximation}

The results presented above have been obtained in the adiabatic
approximation, in which only first derivatives with respect to
time are kept. As stated above, this means that in the iterative
solution of Eq. (\ref{chitild}) only the first iteration has been
maintained. Here we discuss the contribution of the next
iteration. For the sake of simplicity, we carry out this
calculation for an un-biased system with two single-channel leads.

To second order in the temporal derivative, the scattering
solution can be presented in the form
\begin{eqnarray}
\chi({\bf r},t)&=&\chi^{t}({\bf r})\nonumber\\
&-&i\int d{\bf r}'G^{t}(E;{\bf r},{\bf r}')\Bigl
(\dot{\chi}^{t}_{\alpha}({\bf
r}')+\Delta\dot{\chi}^{t}_{\alpha}({\bf r}') \Bigr ),
\end{eqnarray}
with
\begin{eqnarray}
\Delta\dot{\chi}^{t}_{\alpha}({\bf r}')&=&-i\int d{\bf
r}''\frac{d}{dt}\Bigl (G^{t}(E;{\bf r}',{\bf
r}'')\dot{\chi}^{t}_{\alpha}({\bf r}'')\Bigr ).
\end{eqnarray}
It follows that the quantity $U$, Eq. (\ref{u}), is now modified
into
\begin{eqnarray}
\tilde{U}_{\beta\alpha}&=&U_{\beta\alpha}+\Delta U_{\beta\alpha},\nonumber\\
U_{\beta\alpha}&=&\int d{\bf r}\chi^{t}_{\beta}({\bf r})
\dot{\chi}^{t}_{\alpha}({\bf
r}),\nonumber\\
\Delta U_{\beta\alpha}&=&\int d{\bf r}\chi^{t}_{\beta}({\bf
r})\Delta\dot{\chi}^{t}_{\alpha}({\bf r}).
\end{eqnarray}
As a result, we find that the current entering lead $\beta$
consists of two parts, the leading order in the adiabatic
approximation, $I_{\beta}$, which has been discussed above, and a
correction, $\Delta I_{\beta}$. Explicitly,
\begin{eqnarray}
\tilde{I}_{\beta}(t) &=&I_{\beta}(t)+\Delta
I_{\beta}(t),\nonumber\\
I_{\beta}(t)&=&e\int\frac{dE}{2\pi}\sum_{\alpha}f_{\alpha}(E)\nonumber\\
&&\times\Bigl [\delta_{\alpha\beta}-|S^{t}_{\beta\alpha}|^{2}+2\Re
\Bigl
(S^{t\ast}_{\beta\alpha}U_{\beta\alpha}\Bigr )\Bigr ],\nonumber\\
\Delta I_{\beta}(t)&=&e\int
\frac{dE}{2\pi}\sum_{\alpha}f_{\alpha}(E)\nonumber\\
&&\times\Bigl [-|U_{\beta\alpha}|^{2}+2\Re\Bigl
(S^{t\ast}_{\beta\alpha}\Delta U_{\beta\alpha}\Bigr
)\Bigr].\label{deltai}
\end{eqnarray}

Let us first verify that the correction $\Delta I_{\beta}$ obeys
charge conservation over the entire period. To this end, we sum
$\Delta I_{\beta}$ over $\beta$. Using Eq. (\ref{sums}), we have
\begin{eqnarray}
&&\sum_{\beta}\Bigl [|U_{\beta\alpha}|^{2}-2\Re\Bigl
(S^{t\ast}_{\beta\alpha}\Delta U_{\beta\alpha}\Bigr )\Bigr
]\nonumber\\
&=&\int d{\bf r}\int d{\bf r}'\Bigl
[\sum_{\beta}\chi^{t}_{\beta}({\bf r})\dot{\chi}^{t}_{\alpha}({\bf
r})\chi^{t\ast}_{\beta}({\bf
r}')\dot{\chi}^{t\ast}_{\alpha}({\bf r}')\nonumber\\
&-&2\Im\Bigl (\chi^{t\ast}_{\alpha}({\bf
r})\frac{d}{dt}G^{t}(E;{\bf r},{\bf
r}')\dot{\chi}^{t}_{\alpha}({\bf r}')\Bigr )\Bigr ].\label{curcon}
\end{eqnarray}
We can now employ the fact that the current should be conserved
upon integrating over the entire period, that is, when expression
(\ref{curcon}) is inserted into $\oint dt$. Then we may integrate
the second term by parts. Consequently, using the relation
\cite{L00}
\begin{eqnarray}
\sum_{\alpha}\chi^{t}_{\alpha}({\bf r})\chi^{t\ast}_{\alpha}({\bf
r}')=-2\Im G^{t}(E;{\bf r},{\bf r}'),\label{img}
\end{eqnarray}
the two terms in (\ref{curcon}) exactly cancel one another.

Let us now turn to the expression for the correction $\Delta
I_{\beta}$, Eq. (\ref{deltai}), and insert there the condition
that the system is un-biased, i.e., $f_{\alpha}(E)=f(E)$. Then,
using Eqs. (\ref{sums}), (\ref{chitt}), (\ref{dchide}), and
(\ref{img}), we find
\begin{eqnarray}
&&\sum_{\alpha}|U_{\beta\alpha}|^{2}=-2\int d{\bf r}\int d{\bf
r}'\Bigl (\frac{\partial\chi^{t}_{\beta}({\bf r})}{\partial
E}\Bigr )\dot{V}({\bf r},t)\nonumber\\
&&\times\Im\Bigl (G^{t}(E;{\bf r},{\bf r}')\Bigr )\dot{V}({\bf
r}',t)\Bigl (\frac{\partial\chi^{t\ast}_{\beta}({\bf
r}')}{\partial E}\Bigr ),
\end{eqnarray}
and
\begin{eqnarray}
&&2\Re\sum_{\alpha}S^{t\ast}_{\beta\alpha}\Delta U_{\beta\alpha}
=2\Im\int d{\bf r}\int d{\bf r}'\Bigl [\Bigl
(\frac{\partial\chi^{t}_{\beta}({\bf r})}{\partial E}\Bigr )
\dot{V}({\bf
r},t)\nonumber\\
&&\times\Bigl (\frac{\partial G^{t}(E;{\bf r},{\bf r}')}{\partial
E}\Bigr )\dot{V}({\bf r}',t)\chi^{t\ast}_{\beta}({\bf r}')\nonumber\\
&+&\Bigl (\frac{\partial^{2}\chi^{t}_{\beta}({\bf r})}{\partial
E^{2}}\Bigr )\dot{V}({\bf r},t)G^{t}(E;{\bf r},{\bf
r}')\dot{V}({\bf r}',t)\chi^{t\ast}_{\beta}({\bf r}')\Bigr
]\nonumber\\
&+&\Im\int d{\bf r}\Bigl (\frac{\partial^{2}\chi^{t}_{\beta}({\bf
r})}{\partial E^{2}}\Bigr )\ddot{V}({\bf
r},t)\chi^{t\ast}_{\beta}({\bf r}).
\end{eqnarray}
To obtain the last equality, we have made use of the relations
\begin{eqnarray}
\frac{\partial^{2}\chi^{t}_{\beta}({\bf r})}{\partial
E^{2}}=-2\int d{\bf r}'\Bigl (\frac{\partial\chi^{t}_{\beta}({\bf
r}')}{\partial E}\Bigr )G^{t}(E;{\bf r}',{\bf r}),
\end{eqnarray}
and
\begin{eqnarray}
&&\dot{G}^{t}(E;{\bf r},{\bf r}')\nonumber\\
&=&\int d{\bf r}_{1}G^{t}(E;{\bf r},{\bf r}_{1})\dot{V}({\bf
r}_{1},t)G^{t}(E;{\bf r}_{1},{\bf r}').\label{gdot}
\end{eqnarray}
Both relations are obtained by taking derivatives of Eqs.
(\ref{green}) and (\ref{chiteq}) with respect to the energy and
the time.

Collecting all these terms, we obtain
\begin{eqnarray}
&&\Delta I_{\beta }(t)=\frac{-e}{\pi}\int dE
f(E)\nonumber\\
&&\times\Im\frac{\partial}{\partial E}\langle\chi^{t}_{\beta}|
\dot{V}(t)\dot{G}^{t}(E)+\frac{1}{2}\ddot{V}(t)G^{t}(E)|\chi^{t}_{\beta}\rangle.
\label{correction}
\end{eqnarray}
It is satisfactory to note that, again, (upon integrating by parts
with respect to the energy), the energy integral includes the
derivative of the Fermi function. Hence, up to the second-order in
the adiabatic approximation, we find
\begin{eqnarray}
&&\tilde{I}_{\beta}(t)=\frac{e}{2\pi}\int dE\Bigl (\frac{\partial
f(E)}{\partial eE}\Bigr )\Bigl
[\langle\chi^{t}_{\beta}|\dot{V}|\chi^{t}_{\beta}\rangle\nonumber\\
&+&\Im\Bigl (\langle\chi^{t}_{\beta}|2
\dot{V}(t)\dot{G}^{t}(E)+\ddot{V}(t)G^{t}(E)|\chi^{t}_{\beta}\rangle
\Bigr )\Bigr  ].
\end{eqnarray}
To estimate the relative magnitude of the correction (the second
term above) compared to the leading order one (the first term
there), consider the part including the second derivative of the
potential. Using Eq. (\ref{dchide}), this part becomes
\begin{eqnarray}
\int d{\bf r}\chi^{t\ast}_{\beta}({\bf r})\ddot{V}({\bf r},t)\Bigl
(-\frac{\partial\chi^{t}_{\beta}({\bf r})}{\partial E}\Bigr ).
\end{eqnarray}
It follows that the correction will be smaller than the leading
order term by a factor proportional to $1/\tau$, due to the extra
temporal derivative, times the energy derivative of the
instantaneous scattering state. The latter will include the
energy-derivatives of the instantaneous transmission and
reflection amplitudes (which appear in $\chi^{t}$), and possibly a
term proportional to the 1D density of states, i.e., the velocity,
(coming, e.g., from the factors $e^{ika}$ which appear in
$\chi^{t}$ of our simple example discussed in Sec. III). We may
conclude that as long as the energy derivative of the
instantaneous scattering solution is small on the scale of
$\omega\propto 1/\tau$, the leading order adiabatic approximation
suffices to obtain the pumped charge.

\end{multicols}


\begin{references}
\bibitem{AG99}
B. L. Altshuler and L. I. Glazman, Science {\bf 283}, 1864 (1999).
\bibitem{He91}
F. Hekking and Yu. V. Nazarov, Phys. Rev. B {\bf 44}, 9110 (1991).
\bibitem{Br98}
P. W. Brouwer, Phys. Rev. B {\bf 58}, R10135 (1998).
\bibitem{AA98}
I. L. Aleiner and A. V. Andreev, Phys. Rev. Lett. {\bf81}, 1286
(1998).
\bibitem{Z99}
F. Zhou, B. Spivak, and B. L. Altshuler, Phys. Rev. Lett. {\bf
82}, 608 (1999).
\bibitem{SAA00}
T. A. Shutenko, I. L. Aleiner and B. L. Altshuler, Phys. Rev. B
{\bf 61}, 10366 (2000).
\bibitem{S00}
S. H. Simon, Phys. Rev. B {\bf 61}, R16327 (2000).
\bibitem{Ma00}
P. A. Maksym, Phys. Rev. B {\bf 61}, 4727 (2000).
\bibitem{Lev00}
Y. Levinson, O. Entin-Wohlman, and P. W\"{o}lfle, cond-mat/0010494
(2000).
\bibitem{W00}
Y. Wei, J. Wang, and H. Gou, Phys. Rev. B {\bf 62}, 9947 (2000).
\bibitem{Mo01}
M. Moskalets and M. B\"{u}ttiker, cond-mat/0108061 (2001).
\bibitem{Sh01}
P. Sharma and C. Chamon, Phys. Rev. Lett. {\bf 87}, 096401 (2001).
\bibitem{Ma01}
Y. Makhlin and A. Mirlin, Phys. Rev. Lett. {\bf 87}, 276803
(2001).
\bibitem{wei2}
Y. Wei, J. Wang, H. Guo, and C. Roland, Phys. Rev. B {\bf 64},
115321 (2001).
\bibitem{Sw99}
M. Switkes, C. M. Marcus, K. Campman and A. C. Gossard, Science
{\bf 283}, 1905 (1999).
\bibitem{Po92}
H. Pothier, P. Lafarge, U. Urbina, D. Esteve and M. H. Devoret,
Europhys. Lett. {\bf 17}, 249 (1992).
\bibitem{Kou91}
L. P. Kouwenhoven, A. T. Johnson, N. C. van der Vaart, A. van der
Enden, C. J. P. M. Harmans, and C. T. Foxon, Z. Phys. {\bf B85},
381 (1991).
\bibitem{Th83}
D. J. Thouless, Phys. Rev. B {\bf 27}, 6083 (1983).

\bibitem{Bu94}
M. B\"{u}ttiker, H. Thomas, and A. Pr\'{e}tre, Z. Phys. {\bf B94},
133 (1994).
\bibitem{avron}
J. E. Avron, A. Elgart, G. M. Graf, and L. Sadun, Phys. Rev. B
{\bf 62}, R10 618 (2000).
\bibitem{aa}
A. Aharony and O. Entin-Wohlman, cond-mat/0111053 (2001).
\bibitem{L00}
Y. Levinson, Phys. Rev. B {\bf 61}, 4748 (2000).
\bibitem{oew01}
O. Entin-Wohlman, Y. Levinson, and P. W\"{o}lfle, Phys. Rev. B
{\bf 64}, 195308 (2001).

\bibitem{Lev98}
Y. Levinson and P. W\"{o}lfle, Phys. Rev. Lett. {\bf 83}, 1399
(1999).
\bibitem{ng}
T. K. Ng and P. A. Lee, Phys. Rev. Lett. {\bf 61}, 1768 (1988).









\end{references}
\end{document}